\documentclass[epj,twocolumn]{webofc}
\usepackage[varg]{txfonts}   
\usepackage{graphicx}

\wocname{epj}
\woctitle{Seismology of the Sun and the Distant Stars 2016}
\begin{document}
\title{2D dynamics of the radiative core of low mass stars}
%

\author{\firstname{Delphine} \lastname{Hypolite}\inst{1}\fnsep\thanks{\email{delphine.hypolite@cea.fr}} \and
        \firstname{St\'ephane} \lastname{Mathis}\inst{1} \and
        \firstname{Michel} \lastname{Rieutord}\inst{2}
}

\institute{Laboratoire AIM Paris-Saclay, CEA/DRF - CNRS - Universit\'e Paris Diderot, IRFU/SAp Centre de Saclay, F-91191 Gif-sur-Yvette Cedex, France
\and
           Institut de Recherche en Astrophysique et Plan\'etologie, Observatoire Midi-Pyr\'en\'ees, Universit\'e de Toulouse, 14 avenue Edouard Belin, 31400 Toulouse, France
          }

\abstract{%
Understanding the internal rotation of low mass stars all along their evolution is of primary interest
when studying their rotational dynamics, internal mixing and magnetic field generation.
In this context, helio- and asteroseismology probe angular velocity gradients
deep within solar type stars at different evolutionary stages. Still the rotation close to the center of such stars on the
main sequence is hardly detectable and the dynamical interaction of the radiative
core with the surface convective envelope is not well understood. For instance, the
influence of the differential rotation profile sustained by convection and applied
as a boundary condition to the radiation zone is very important in
the formation of tachoclines. 
In this work, we study a 2D hydrodynamical model of a radiative core when an imposed, solar or anti-solar, differential rotation is applied at the upper boundary. This model uses the Boussinesq approximation and we find that the shear induces a cylindrical differential rotation associated with a unique cell of meridional circulation in each hemisphere (counterclockwise when the shear is solar-like and clockwise when it is anti-solar).
The results are discussed in the framework of seismic observables (internal rotation rate, core-to-surface rotation ratio)
while perspectives to improve our modeling by including magnetic field or
transport by internal gravity waves will be discussed.
}
\maketitle
%
\section{Introduction}
\label{intro}

Asteroseismology has revealed the internal rotation profile of numerous stars over the past few years. On one hand, we now have partial access to the dynamical state of low mass stars on the main sequence and during sub giant and red giant evolution phases (\cite{beck12}, \cite{deheuvels12}, \cite{deheuvels14}, \cite{deheuvels16}, \cite{benomar15}). 
On the other hand, the internal rotation profile of the Sun deep within the star, until $0.2R_\odot$, has been inverted (\cite{couvidat03}, \cite{garcia07}). Both helioseismology and asteroseismology show that a strong transport of angular momentum occurs in radiative zones.
These constrains call for a new generation of stellar models since current 1D stellar models do not recover inverted quantities such as for instance the flat rotation profile of the radiative zone of the Sun (e.g. \cite{TC98}, \cite{TC10}) and of more evolved stars (\cite{eggenberger12}, \cite{ceillier13}, \cite{marques13}).

Indeed, one who wants to model the evolution of stars gets immediately confronted with the complexity of stellar rotation (\cite{meynetmaeder00}, \cite{maeder09}).
It is known to have multiple strong consequences on stellar evolution since it drives rotational mixing and transport of angular momentum through the stable radiative zones (\cite{zahn92}, \cite{maederzahn98}, \cite{mathiszahn04}) on secular timescale. In addition, it is connected with the generation of stellar magnetic fields and the presence of tachoclines (\cite{spiegelzahn92}, \cite{GM98}, \cite{brun06}, \cite{strugarek11}) requesting a deep understanding of the stellar rotation.
One of the most challenging aspects is the multidimensional nature of rotating stars (\cite{MR06}, \cite{ELR07}).
Indeed, the centrifugal acceleration leads to the rise of baroclinic flows which can not be fully described by low spherical harmonics expansion and/or latitudinal averages.
For these reasons, we propose a 2D approach to describe the secular evolution of the internal dynamics of rotating radiation zones.
In this framework, the convective envelope applies a latitudinal shear through differential rotation on the underlying radiative core in low mass stars. For example, in the Sun, the rotation profile of the envelope is conical with fast equatorial regions and slower poles (\cite{schou98}, \cite{brun02}).
Past studies have looked for the consequences of such a shear in the solar case (\cite{friedlander76},\cite{garaud02}). However, it calls for a generalisation of the dynamical boundary conditions since 3D numerical simulations reveal that solar-like differential rotation is a function of stellar fundamental parameters and mean rotation rate while anti-solar differential rotation can also be expected in cool stars (\cite{matt11}, \cite{kapyla14}, \cite{gastine14}, \cite{varela16}).

In this framework, the question we address is: \textit{How does the differential rotation of a convective envelope affect the dynamics (differential rotation and meridional circulation) of the radiative core in low mass stars?}

In this work, we use 2D Boussinesq modeling to perform a parametric study over the shear applied as a boundary condition on the radiative core and compute the latitudinally averaged core to the surface rotation ratio of the resulting models which can be compared with seismic observables.

We describe the 2D modeling of the radiative core of low mass stars and its main assumptions in Sect.~\ref{sec:1}.
In Sect.~\ref{sec:2}, we compute the core-to-surface rotation ratio numerically as a function of the shear applied at the top of the domain and discuss the fast rotating solar case. We give our conclusions and perspectives in Sect.~\ref{sec:3}.

\section{Modeling low mass stars radiative core in 2D}
\label{sec:1}

We consider a viscous flow enclosed within a rotating spherical shell.
We suppose the centrifugal acceleration does not deform it. 
We study the interaction of the combined effect of the inertia with rotation in a stably stratified environment, leaving aside magnetic fields (\cite{garaud02b}, \cite{mathiszahn05}, \cite{strugarek11}), internal gravity waves (\cite{TC05}, \cite{TC05b}) and anisotropic turbulence (\cite{zahn92}, \cite{mathis04}, \cite{KB12}) as a first step.
Because we are interested in the evolution of the fluid on secular timescale, we solve as a first step the steady equation of the vorticity combined with the equations of energy and continuity using the Boussinesq approximation (\cite{MR06}, \cite{RB14}, \cite{HR14}).
This approximation retains density variations only in the buoyancy term taking into account both the gravity and centrifugal accelerations and implies that the stratification is only taken into account through the Brunt-V\"ais\"al\"a frequency profile (positive in a radiative zone).
It is given as an input of the simulations using the 1D MESA stellar evolution code (\cite{paxton10}) to generate ZAMS low mass stars radiative core models (we use a metallicity of $Z=0.02$, and a mixing length theory parameter of $\alpha_{\rm MLT}=2$) and we seek the resulting velocity field, i.e. the differential rotation and the meridional circulation for a $1M_\odot$ model.
As developed by \cite{MR06}, hereafter R06, and \cite{HMR16b}, the scaled equations we solve are 
\begin{multline}
\left \{
\begin{array}{lcl}
\vec{\nabla} \times (\vec{e}_z \times \vec{u} - \theta_T \vec{r} - E\Delta\vec{u}) = - n^2(r) \sin{\theta}\cos{\theta}\vec{e}_{\varphi}\; ,\\
\left(\frac{n^2(r)}{r} \right)u_r = \frac{E}{Pr}\Delta\theta_T\; ,\\
\vec{\nabla}\cdot\vec{u} = 0\; ,\\
\end{array}
\right.
\label{eq7}
\end{multline}
where $\vec{u}$ is the scaled velocity field in the corotating frame, $\theta_T$ is the scaled temperature profile and $n$ is the scaled Brunt-V\"ais\"al\"a frequency profile.  
We remove dimensions using $R_c$, the radius of the radiative core, as a length scale, $V$, the baroclinic velocity as a velocity scale, $T'$ as a temperature scale and $\mathcal{N}$ the scale of the Brunt-V\"ais\"al\"a frequency profile both from R06 (We refer the reader to R06 for the detailed description of the adimensionalization).
 
The physical parameters that characterize the system are the Ekman number $E$ which quantifies the importance of the viscosity (we note the kinematic viscosity $\nu$) over the Coriolis effects 
\begin{equation}
E=\frac{\nu}{2\Omega_0R_c^2}\; ,
\end{equation}
where $\Omega_0$ is the rotation rate of the corotating frame, 
and the Prandtl number $Pr$ which is the ratio of the kinematic viscosity over the thermal diffusivity $\kappa$
\begin{equation}
Pr=\frac{\nu}{\kappa}\; .
\end{equation}
The Ekman number is evaluated and found to be very small in radiative core of low mass stars (around $10^{-11}$). We use here the smallest value that numerical resolution allows, i.e. $E=10^{-6}$. Finally, we use a solar value for the Prandtl number, namely $Pr=2.10^{-6}$ (\cite{brun06}).

In this framework, 3D simulations (\cite{matt11}, \cite{kapyla14}, \cite{gastine14}, \cite{varela16}) found that the differential rotation is conical in convective envelopes of low mass stars. We therefore impose at the top of the radiative core
\begin{equation}
\Omega_{cz}(r=R_c,\theta)= \Omega_0 + \Delta\Omega\sin^2\theta\; .
\end{equation}
In the corotating frame, the azimuthal dimensionless velocity reads
\begin{equation}
u_\varphi(r=1,\theta)= b \sin^3\theta\; ,
\label{eq:1}
\end{equation}
where the dimensionless number $\displaystyle{b=\frac{R_c \Delta\Omega}{V}}$ quantifies the shear at the top of the radiative core. The differential rotation is solar-like when the equatorial regions rotate faster than the pole ($b>0$) and anti-solar when they rotate slower ($b<0$). 

At the radiative-convective envelope, the meridional components of the velocity are set to zero illustrating the non penetrative case $u_r=0$ and assuming that the presence of a convective envelope induces $u_\theta=0$ in the corotating frame.

\begin{figure}
\center{
	\resizebox{1.1\columnwidth}{!}{
	\includegraphics{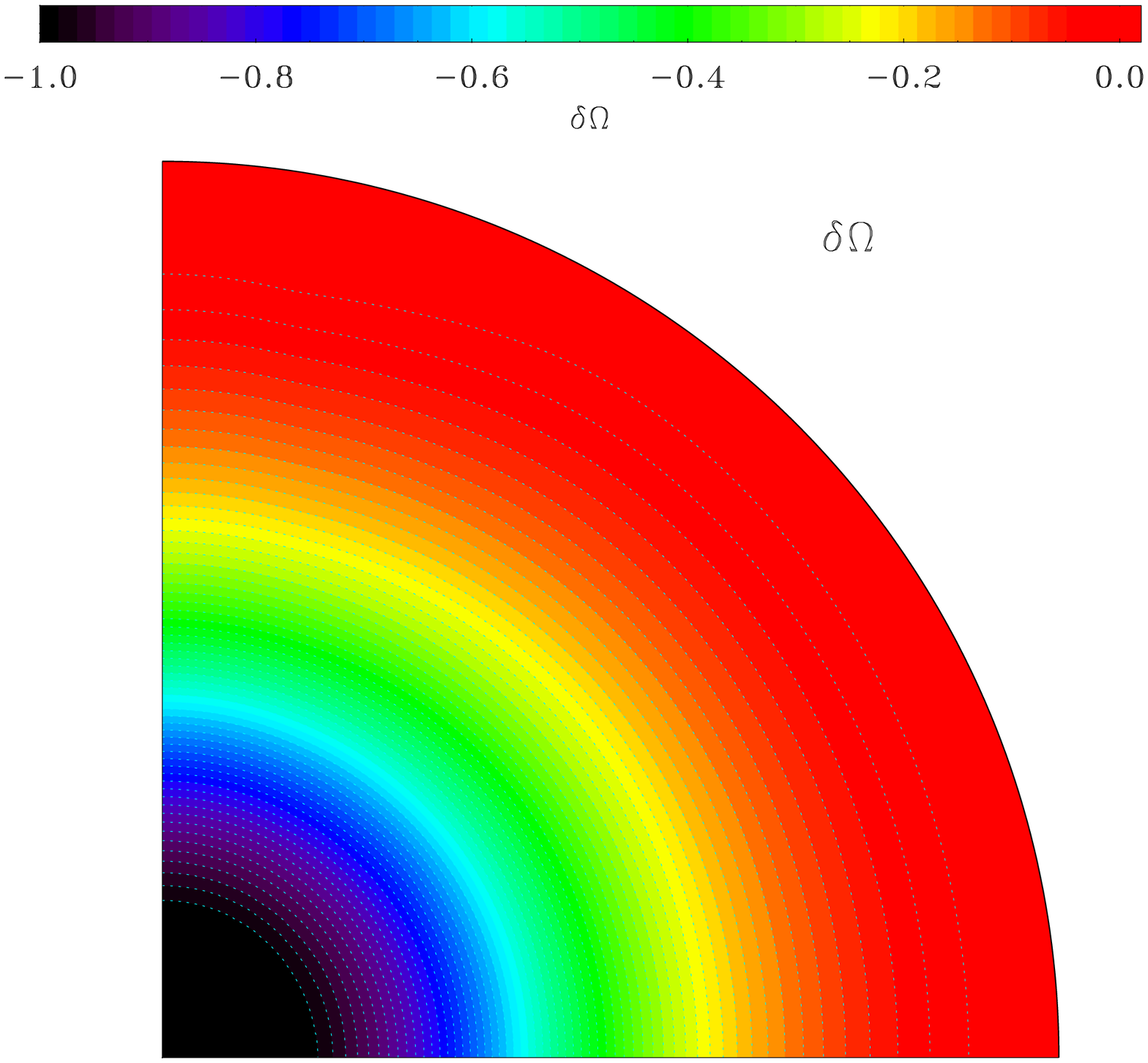}	
	\includegraphics{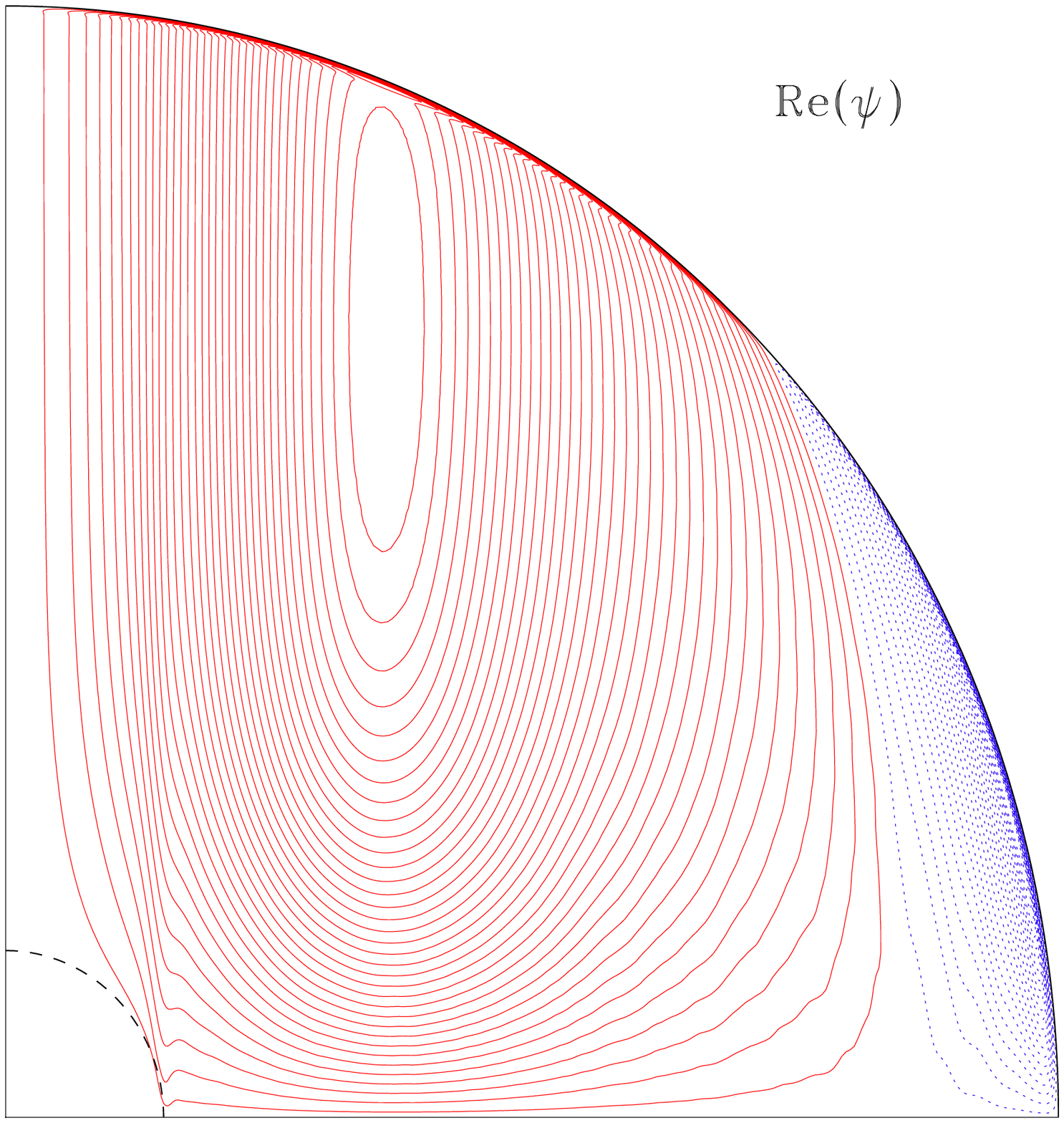}}
	\caption{Differential rotation $\delta \Omega$ and meridional circulation stream function $\psi$ (red : direct sens, blue: clockwise sens)  shown in the meridional plane for an Ekman number $E=10^{-6}$, a Prandtl number $\Pr=2.10^{-6}$ and $b=10^{-2}$. This is the pure baroclinic dynamics.  The stellar rotation axis is vertical.}
	\label{fig:3}}
\end{figure}

The shear we impose by this dynamical boundary condition generates a geostrophic flow, i.e. a flow respecting the geostrophic balance where the pressure gradient compensates the Coriolis acceleration,  characterized by a columnar differential rotation which depends only on the distance to the rotation axis as predicted by the Taylor-Proudman theorem.
Without any shear or for a very weak shear $|b|<10^{-2}$, the flow is driven by the baroclinic torque $-n^2(r)\sin\theta\cos\theta\vec{e}_\varphi$ in the vorticity equation. The solution generated and described by R06 in Eq. (7) is the thermal wind characterized here by a quasi shellular differential rotation as shown in Fig. \ref{fig:3} (left panel). The difference between the flow obtained by R06 and the one we show in Fig. \ref{fig:3} comes from the fact that the Brunt-V\"ais\"al\"a frequency profile used is different in R06 (polynomial interpolation) than the one we use (input of a $1M_\odot$ ZAMS MESA model) and the boundary conditions are stress-free in R06. 

\begin{figure*}
\center{
	\resizebox{2.2\columnwidth}{!}{
	\includegraphics{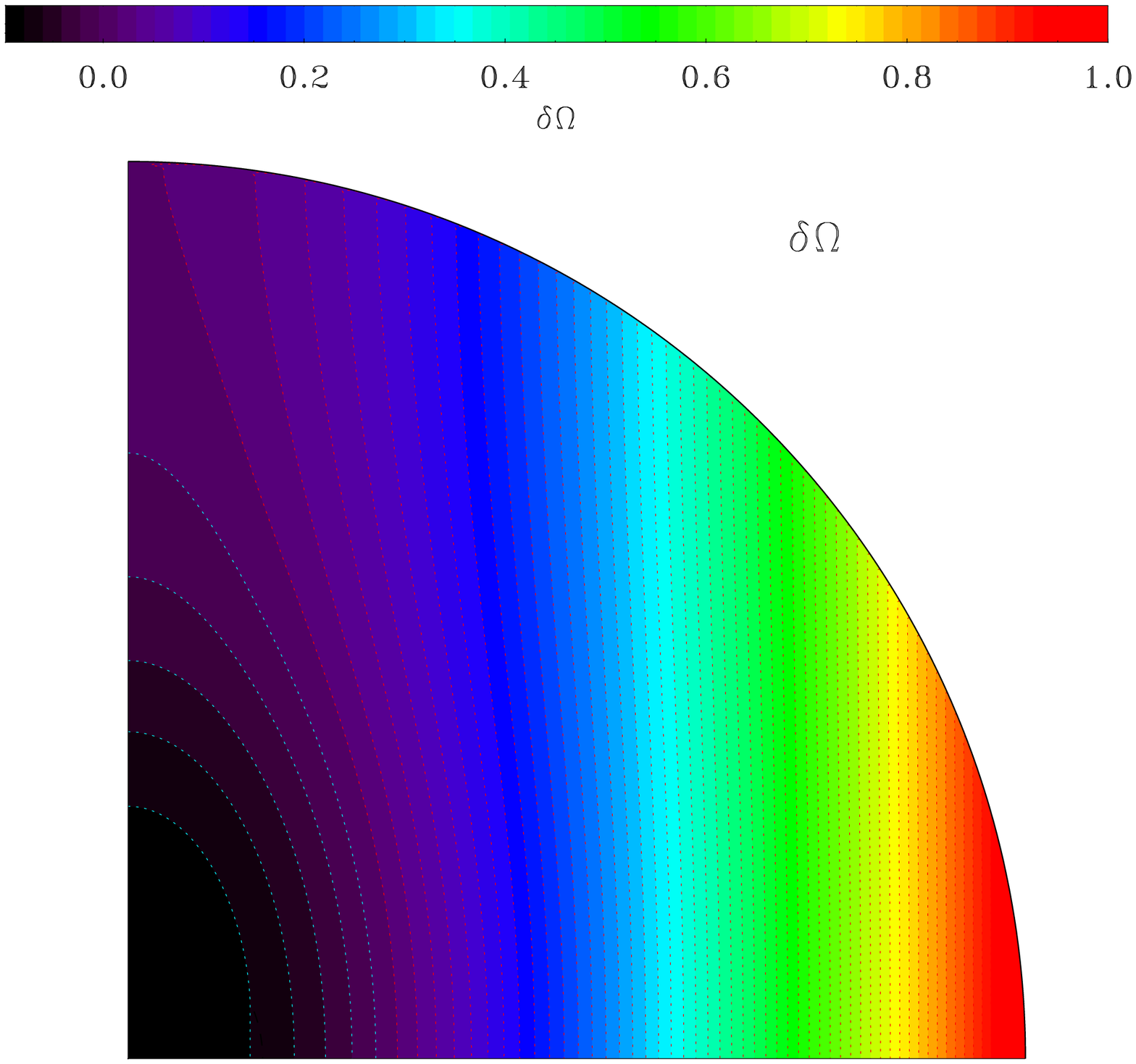}	
	\includegraphics{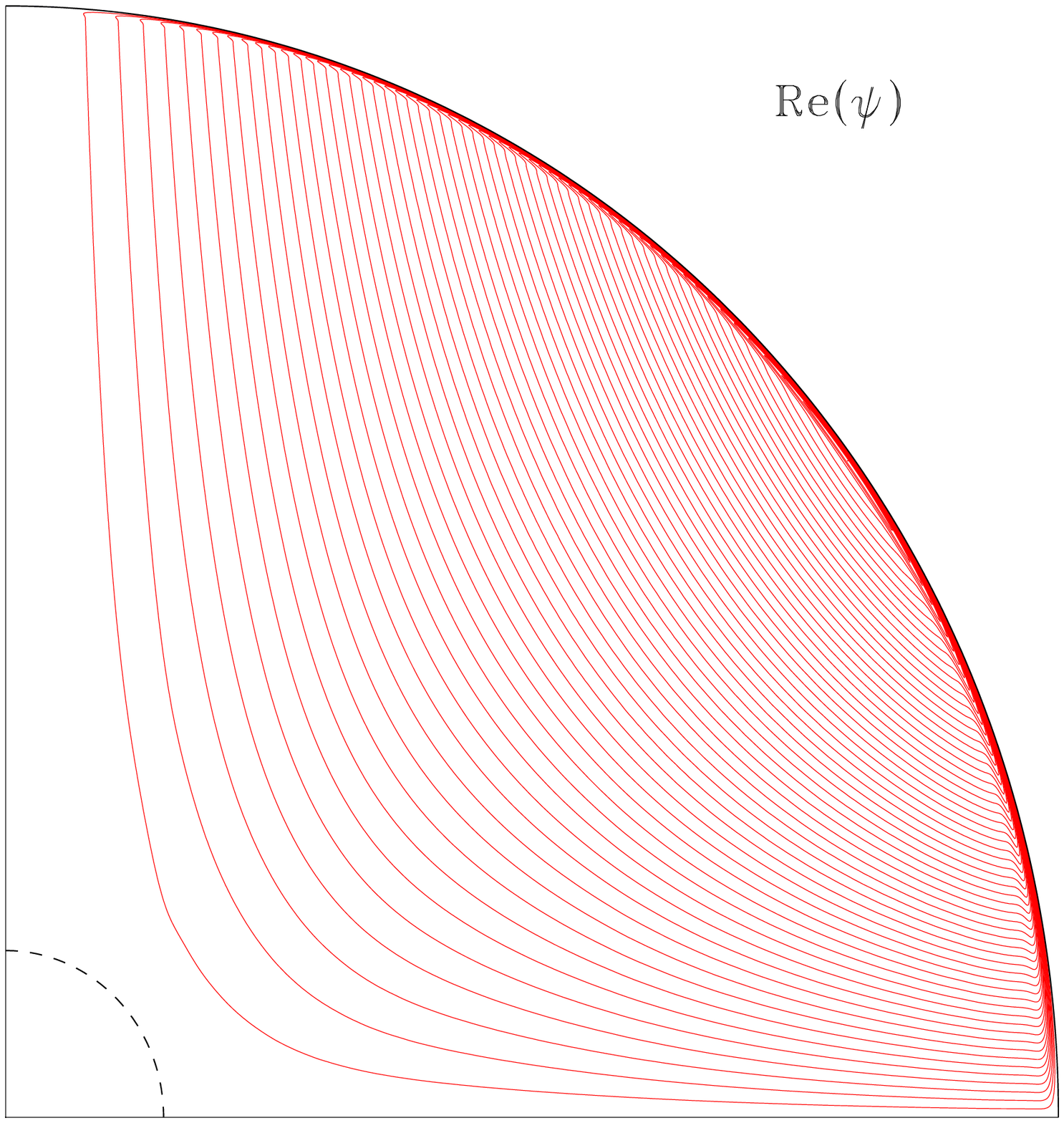}
	\includegraphics{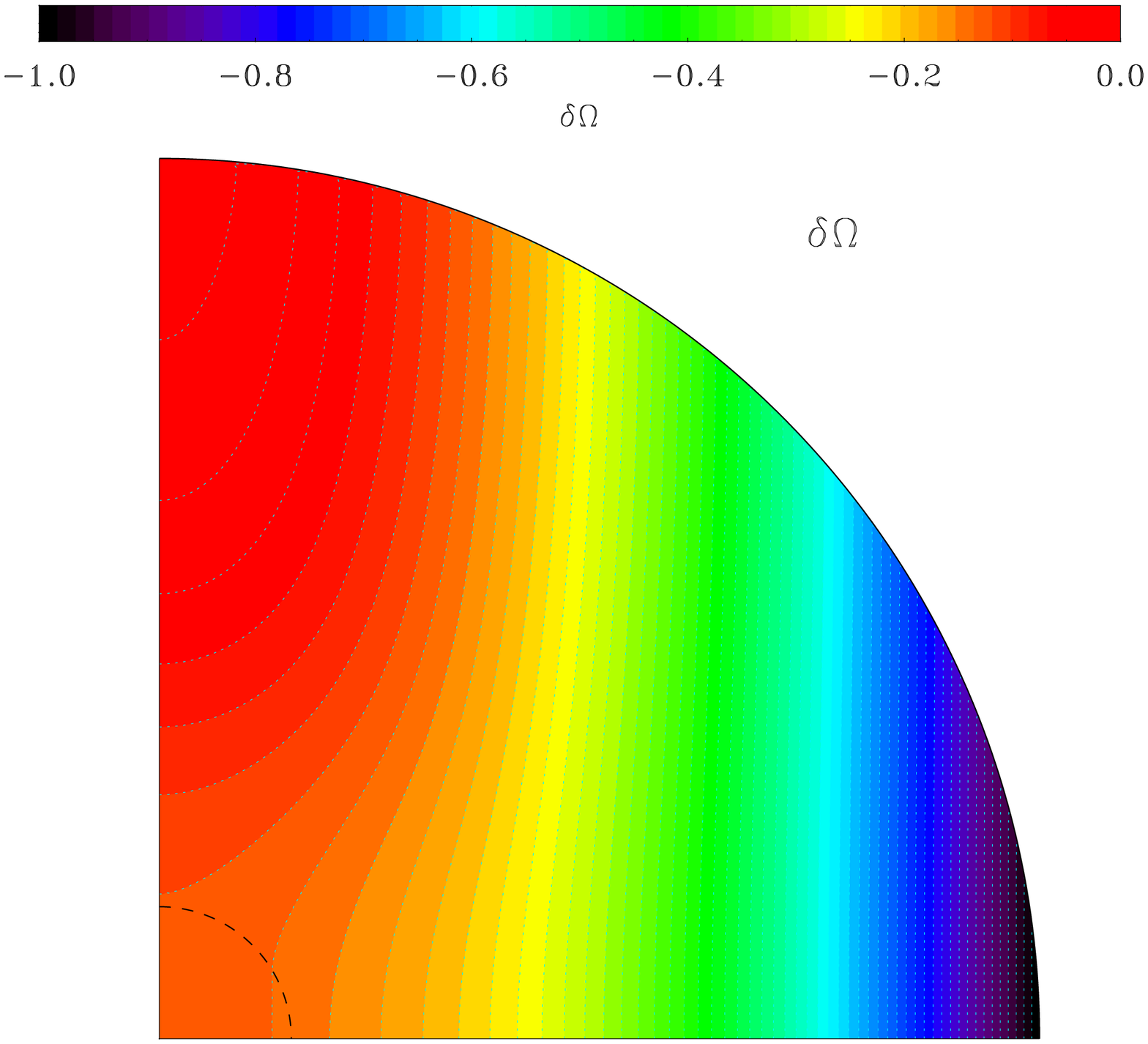}	
	\includegraphics{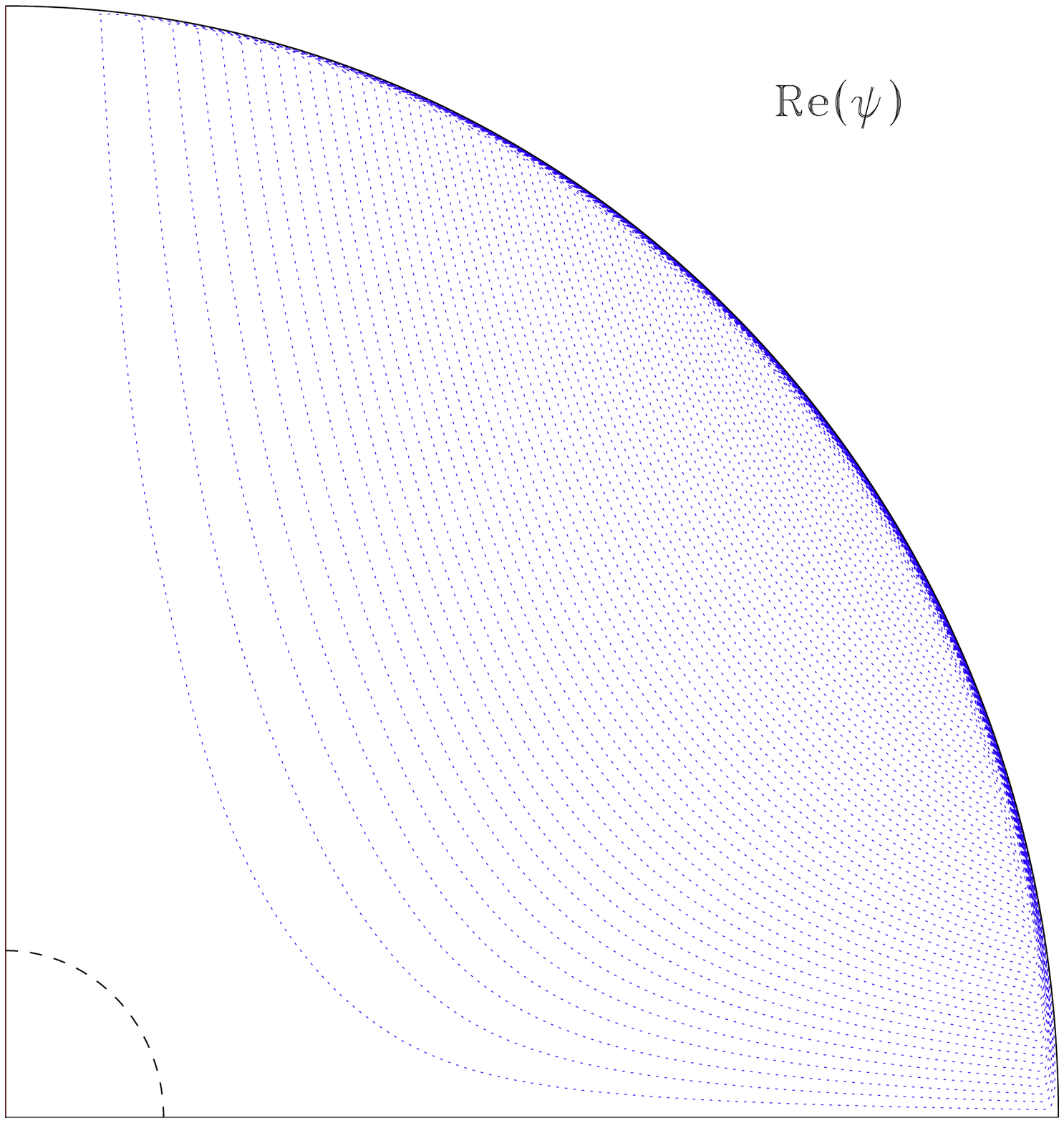}}}
	\caption{Differential rotation $\delta \Omega$ and meridional circulation stream function $\psi$ (red : direct sens, blue: clockwise sens)  shown in the meridional plane for an Ekman number $E=10^{-6}$, a Prandtl number $\Pr=2.10^{-6}$ with $b=10$ (solar like rotation) on the left and $b=-10$ (anti-solar rotation) on the right.  The stellar rotation axis is vertical.}
	\label{fig:1}
\end{figure*}

When $|b|>10^{-2}$, the amplitude of the geostrophic flow arising from the shear applied at the upper boundary is larger than the baroclinic solution, which would dominate otherwise, and the Taylor-Proudman balance tends to be restored leading to a quasi-columnar structure.
In Fig. \ref{fig:1}, we show the resulting differential rotation $\delta\Omega$ for $b=10$ and $b=-10$. The field is normalized by its extremum value and we subtract the rotation of the pole so it is zero there and the differential rotation in the bulk is relative to it.
Therefore, the differential rotation tends towards a cylindrical profile which is very different from the shellular profile assumed by 1D stellar models. 
The associated meridional circulation is dominated by a single, global circulation cell in each hemisphere.
For $b$ positive (solar-like differential rotation), the meridional circulation is counter clockwise.
At the surface of the model of the radiative core, the fluid moves toward the pole which rotates slower than the equator.
Conversely, for negative $b$, the meridional circulation is clockwise and towards the equatorial regions at the radiative-convective interface.

\section{Seismic observables}
\label{sec:2}

\begin{figure}
\center{
\resizebox{1\columnwidth}{!}{
  \includegraphics{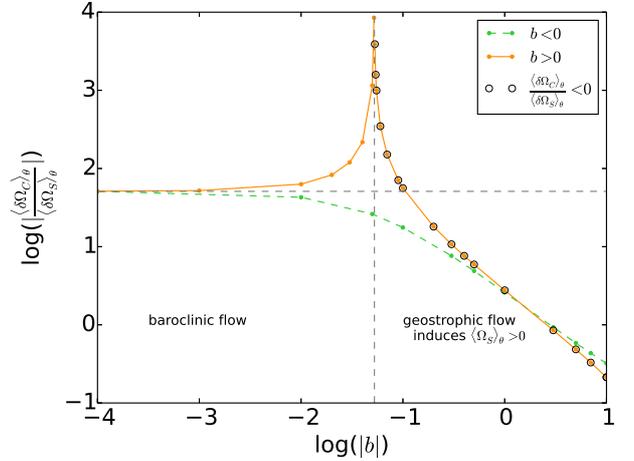}}}
	\caption{Logarithm of the averaged core-to-surface rotation ratio as a function of the logarithm of the shear parameter $b$ for an Ekman number $E=10^{-6}$ and a Prandtl number $\Pr=2.10^{-6}$. 
	Orange dots are for positive $b$, green dots for negative $b$. The circled dots are negative core-to-surface rotation ratio (non circled are positive).
When $b\ll 1$ the dynamics is dominated by the baroclinic flow and the core-to-surface rotation ratio is constant (horizontal dashed line). For $b$ positive, $\langle\delta\Omega_S\rangle_\theta$ changes sign inducing a singularity located at the vertical dashed line.
		 \label{fig:2a} }  
\end{figure}

The core-to-surface rotation ratio is the latitudinally averaged rotation rate at the top of the radiative core divided by the latitudinally averaged central rotation rate $\langle\delta\Omega_C\rangle_\theta/\langle\delta\Omega_S\rangle_\theta$ where $\langle \ldots\rangle_\theta=\int_0^{\pi/2}\ldots \sin\theta d\theta$. 
This ratio has been inverted in asteroseismology on several low mass stars.
This only seismic diagnosis as of today is computed to characterize its behavior according to the shear strength we impose.

In our set-up, at the radiative-convective interface, an Ekman layer arises due to rotation and viscosity.
This layer permits the velocity field (the sum of both the baroclinic flow and the geostrophic one) to match the boundary conditions by adding corrections whose amplitudes are important within the layer and weak outside of it.
The Ekman layer has a characteristic thickness of $\sqrt{E}$, where we recall that $E$ is the dimensionless Ekman number. We therefore compute the averaged surface rotation rate $\langle\delta\Omega_S\rangle_\theta$ just outside of it to avoid the boundary layer corrections. Moreover, we do not compute it at $r=1$ since it would tend toward zero for small $b$ and consequently the ratio $\langle\delta\Omega_C\rangle_\theta/\langle\delta\Omega_S\rangle_\theta$ would diverge.
The averaged core rotation rate $\langle\delta\Omega_C\rangle_\theta$ is computed at the radius $r=0.15 R_c$ because g-modes, that would allow to probe central regions, are hardly identifiable closer to the center (\cite{appourchaux10}, \cite{garcia10}).

We display the ratio $\langle\delta\Omega_C\rangle_\theta/\langle\delta\Omega_S\rangle_\theta$ as a function of $b$ in Fig. \ref{fig:2a} in logarithmic values.
When $|b|$ tends towards zero, the ratio tends to a finite value.
This is because, when the shear is very small, the dynamics is dominated by the baroclinic flow described by R06 visible in Fig. \ref{fig:3}. With the Brunt-V\"ais\"al\"a frequency profile from MESA, the differential rotation is shellular and since we subtract the rotation of the pole, the differential rotation is negative in the entire zone. Therefore the ratio is positive that corresponds to negative averaged core rotation rate divided by a negative averaged surface rotation rate.

When $b$ is negative, the ratio stays positive even when the shear becomes important and the geostrophic flow dominates. Moreover, we observe that when the absolute value of the shear increases, the ratio decreases. Indeed, the shear implies a transport of angular momentum, proportional to $b$ both in the radial and latitudinal directions, which tends to accelerate the superficial regions more than the central regions.

If $b$ is positive, the ratio displays a singularity around $b\sim0.053$.
This singularity is due to the change of sign of $\langle\delta\Omega_S\rangle_\theta$.
Indeed for $b\geq0.053$, the acceleration of the equator we impose through the shear boundary condition induces a positive $\langle\delta\Omega_S\rangle_\theta$.
The ratio is then negative and has the same behavior than the branch of negative $b$. Namely, when $b$ increases, the absolute value of the ratio decreases due to the enhanced transport of angular momentum.
Roughly on the right side of the singularity, the solution is dominated by the geostrophic flow.

The position of the singularity depends on the zero we choose as a reference for the differential rotation; here, we recall that we choose the rotation of the pole to be zero.
It also depends on the radius at which we compute $\langle\delta\Omega_S\rangle_\theta$. Indeed, if we compute it at $r=1$, the ratio $\langle\delta\Omega_C\rangle_\theta/\langle\delta\Omega_S\rangle_\theta$ would diverge for small $|b|$ since the boundary conditions would reduce to no-slip ones.
The singularity would disappear as it would coincide with the divergence at small $b$ and the sign of the ratio $\langle\delta\Omega_C\rangle_\theta/\langle\delta\Omega_S\rangle_\theta$ would be the opposite of the sign of $b$ for all $b$.

Since we work in the corotating frame at an unknown rotation rate and we set the rotation of the pole to be zero, this study provides only a qualitative behavior of the core-to-surface rotation ratio and the sign has a relative physical meaning.
The important behavior to remember is the decrease of the ratio as the shear is strong.
 
\begin{figure}
\center{
  \resizebox{1\columnwidth}{!}{
  \includegraphics{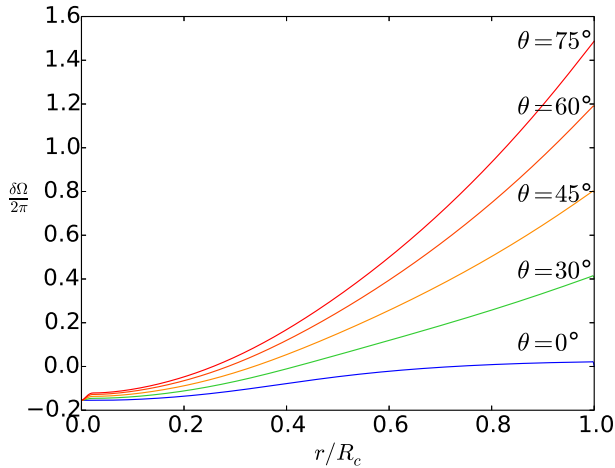}}}
	\caption{Relative radial differential rotation profiles within the radiative zone for $b=10$ and $E=10^{-6}$ at different colatitudes. \label{fig:2b} }
\end{figure}

In the derivation of the Eqs. (\ref{eq7}), we assumed that the Brunt-V\"ais\"al\"a frequency $N^2$ is close to $4\Omega^2$ which is the fast rotating case where the transient baroclinic timescale is short enough for the steady state study to be relevant (R06, \cite{Rbouquinconf}).
This means that a fast rotating solar case (a young Sun) is reachable for $b=10$.
For this value, the averaged core-to-surface rotation rate ratio is expected to be rather small, less than one in agreement with observations (\cite{benomar15}). But the internal rotation profile is far from flat. As we can see in Fig. \ref{fig:2b}, the rotation profiles for different colatitudes within the radiative core do not coincide as observed by heliosismology.
Moreover, we do not produce a tachocline, pointing out that our current modeling is lacking a physical process responsible for the current rotation field of the solar radiative core (for reviews, see \cite{mathis13}, \cite{brun15}). 
\section{Conclusions}
\label{sec:3}

We build a 2D model of fast rotating radiative cores in low mass stars undergoing the shear of a differentially rotating convective envelope.

This shear is responsible for a redistribution of angular momentum deep within the star and gives rise to a geostrophic solution with a cylindrical differential rotation.
Such a profile is very different from the shellular rotation profile assumed in 1D models.
These models should therefore be used carefully in the radiative core of low mass stars when the shear applied by a convective envelope is important, i.e. $b>10^{-2}$.
Regarding the seismic observables such as the averaged core-to-surface rotation rate or the internal rotation profile, we conclude that we have to take into account other processes responsible for efficient transport of angular momentum both in the radial and in the latitudinal directions.
Such processes may be internal gravity waves (\cite{zahn97}), magnetic fields (\cite{GM98}, \cite{spruit99}, \cite{strugarek11}, \cite{AGW13}) or an anisotropic turbulence (\cite{zahn92}, \cite{maeder03}, \cite{mathis04}).
These candidates are considered for implementation in our current 2D modeling of radiative zones. 

\subsection*{Acknowledgments}
D.H. and S.M. acknowledge funding by the European Research Council through ERC grant SPIRE 647383.
The authors acknowledge funding by SpaceInn and PNPS (CNRS/INSU) and by CNES CoRoT/PLATO grant at SAp and IRAP.

\end{document}